\documentclass[12pt]{iopart}

\usepackage{graphicx}
\usepackage{epsfig}
  
\begin{document}

\article[]{Topical review}{High-field $\mu$SR studies of 
superconducting and magnetic correlations 
in cuprates above $T_c$}

\author{J E Sonier}

\address{Department of Physics, Simon Fraser University,
Burnaby, British Columbia V5A 1S6, Canada}
\ead{jsonier@sfu.ca}
\begin{abstract}
The advent of high transverse-field muon spin rotation (TF-$\mu$SR) has 
led to recent $\mu$SR investigations of the magnetic-field response of cuprates
above the superconducting transition temperature $T_c$. Here the
results of such experiments on hole-doped cuprates are reviewed. 
Although these investigations are currently ongoing, it is clear that
the effects of high field on the internal magnetic field distribution
of these materials is dependent upon a competition between superconductivity 
and magnetism. In La$_{2-x}$Sr$_x$CuO$_4$ the response to the external field 
above $T_c$ is dominated by heterogeneous spin magnetism. However, the magnetism
that dominates the observed inhomogeneous line broadening below $x \! \sim \! 0.19$ is 
overwhelmed by the emergence of a completely different kind of magnetism 
in the heavily overdoped regime. The origin of the magnetism above $x \! \sim \! 0.19$
is currently unknown, but its presence hints at a competition between 
superconductivity and magnetism that is reminiscent of the underdoped regime.
In contrast, the width of the internal field distribution of underdoped 
YBa$_2$Cu$_3$O$_y$ above $T_c$ is observed to track $T_c$ and the density 
of superconducting carriers. This observation suggests that the
magnetic response above $T_c$ is not dominated by electronic moments, but
rather inhomogeneous fluctuating superconductivity. 
\end{abstract}


\maketitle

\section{Introduction}
An applied magnetic field has been a revealing tuning parameter in the
study of high-$T_c$ cuprates. Superconducting properties can be probed 
via the diamagnetic response of these materials to an external 
magnetic field. With TF-$\mu$SR, information on the density of the superconducting 
carriers can be obtained by measuring the internal magnetic field 
distribution of a lattice of vortices generated by an external field \cite{Sonier:00}. 
An applied magnetic
field can also be used to partially or fully suppress superconductivity 
at low temperature $T$, to reveal competing ground states \cite{Sachdev:03} 
and to detect normal quasiparticles 
\cite{Boebinger:96,Dorion-Leyraud:07,Yelland:08,Vignolle:08}.
The high field required to reach the normal state of cuprates at low $T$ 
is usually unattainable in a TF-$\mu$SR experiment. Consequently, TF-$\mu$SR
studies at low $T$ are restricted to detecting the effects of field-induced
suppression of superconductivity in and around the
vortex cores \cite{SonierRPP:07}.

In recent years high TF-$\mu$SR has found an application in the study of the
magnetic response of cuprates in the normal state above $T_c$. It has long 
been believed that this region of the cuprate phase diagram
holds vital clues to understanding the 
microscopic mechanism of high-$T_c$ superconductivity. For example, 
cuprates exhibit a pseudogap at the Fermi surface above $T_c$ \cite{Timusk:99}. 
There is still no consensus on the origin of the pseudogap, but establishing
agreement on the specific question of whether it is associated with preformed
Cooper pairs (paired electrons) or a competing phase would be 
a major advancement in the field. The magnetic response of cuprates 
above $T_c$ became a topic of immense interest after 
Ong \etal \cite{Xu:00,Wang:01,Wang:06} demonstrated that a sizeable magnetic 
field creates a Nernst effect persisting far above $T_c$. 
The Nernst signal is accompanied by field-enhanced 
diamagnetism that persists in very large applied magnetic fields \cite{Li:05,Wang:05}. 
This behaviour is qualitatively distinct from the fluctuating diamagnetism 
observed in conventional BCS superconductors \cite{BCS:57}, which is rapidly 
quenched in high fields. 

The Nernst measurements accompany a growing body of experiments on high-$T_c$ 
cuprates that hint at superconducting pair 
correlations surviving on short time and length scales far above $T_c$.
With increasing temperature, superconductivity is ultimately 
destroyed by fluctuations of the pairing amplitude, fluctuations of the phase, 
or both. In a conventional BCS superconductor
the typical energy scale of phase fluctuations
is much larger than the energy gap, which in turn is proportional to
$T_c$. Consequently, it is the vanishing of the pairing amplitude and not
phase fluctuations that determines $T_c$. Nevertheless, just above $T_c$ there
is a critical region characterized by amplitude (Gaussian) fluctuations of 
short-lived Cooper pairs. The superconducting fluctuations in this region
give rise to weak diamagnetism, excess specific heat, conductivity and Josephson 
tunneling \cite{Skocpol:75}.  

The situation has been argued to be different in the high-$T_c$ cuprates 
\cite{Emery:95}, because phase fluctuations are 
enhanced by the reduced dimensionality associated with superconductivity 
being more or less confined to the CuO$_2$ layers.
In addition, a low density of superconducting carriers in the underdoped regime 
results in a reduced stiffness to phase fluctuations. These factors favour 
a situation where upon cooling, bulk superconductivity is established by the onset of 
{\it long-range} phase coherence at $T_c$. Consequently, the simple binding of 
electrons into Cooper pairs may occur at temperatures well above $T_c$ where amplitude 
fluctuations are still relatively weak.    
High-frequency conductivity measurements by Corson \etal
\cite{Corson:99} show a residual Meissner effect in the normal state of
Bi$_2$Sr$_2$CaCu$_2$O$_{8 + \delta}$ associated with a finite phase stiffness 
on short time scales. Furthermore, enhanced fluctuating diamagnetism that 
is uncharacteristic of conventional amplitude fluctuations has been observed 
in several cuprates above $T_c$
\cite{Kanoda:88,Carballeira:00,Carretta:00,Lascialfari:02,Lascialfari:03,Cabo:06}.
      
The above results on cuprates have been explained in terms of a  
Kosterlitz-Thouless transition, below which long-range superconducting 
order occurs. In this model the Nernst effect is attributed to 
the destruction of long-range phase coherence of the Abrikosov vortex lattice
({\it i.e.} creation of a vortex liquid) in a two-dimensional system 
by thermally generated mobile vortices above the Kosterlitz-Thouless transition. 
This interpretation implies that there must be short-range 
phase coherence among Cooper pairs in sizeable regions of the sample 
in order to nucleate vortices. 
Yet the Nernst signal above $T_c$ has also been attributed
to amplitude fluctuations of short-lived Cooper pairs \cite{Ussishkin:02}.
Pourret \etal \cite{Pourret:06} have shown that a Nernst signal caused by 
amplitude fluctuations is observable in a dirty BCS superconductor far 
above $T_c$. Likewise, Rullier \etal \cite{Rullier:06} 
have demonstrated that the Nernst 
region for YBa$_2$Cu$_3$O$_y$ is expanded by disorder above a suppressed value of $T_c$ 
\cite{Rullier:06}, and Bergeal \etal \cite{Bergeal:08} have provided
evidence for amplitude fluctuations in cuprates above $T_c$ from measurements
of the Josephson effect in an optimally doped/underdoped junction.
It is thus unclear from the experiments done thus far on cuprates
whether the region above $T_c$ is dominated by amplitude 
or phase fluctuations.

There is also growing evidence that the superconducting correlations
in cuprates first develop in nanoscale regions at temperatures well above 
$T_c$ \cite{Alvarez:01,Kresin:06,Mello:07}.
Indeed, Gomes \etal \cite{Gomes:07} have observed the development of spatially 
inhomogeneous pairing gaps above $T_c$ via scanning tunnelling 
microscopy (STM) on Bi$_2$Sr$_2$CaCu$_2$O$_{8 + \delta}$. With decreasing temperature
the gapped regions proliferate in a way that is consistent with the eventual
formation of the bulk superconducting phase below $T_c$ via percolation or 
Josephson coupling of these regions. 
The observation of hysteresis in low-field
magnetization measurements on La$_{2-x}$Sr$_x$CuO$_4$ 
by Panagopoulos \etal \cite{Panagopoulos:06}
could also be interpreted as indirect evidence for fluctuating 
diamagnetic superconducting domains above $T_c$. 
More recently Kanigel \etal \cite{Kanigel:08} observed
a Bogoliubov-like electronic dispersion in the pseudogap 
region of Bi$_2$Sr$_2$CaCu$_2$O$_{8 + \delta}$, 
also suggestive of short-range superconducting order.  

New studies of the Nernst signal in cuprates \cite{Cyr:09,Daou:09}
have now clearly distinguished the individual contributions from quasiparticles and
superconducting fluctuations (phase or amplitude). The onset of a temperature 
dependence of the quasiparticle Nernst signal follows the doping dependence of the
pseudogap temperature $T^*$ and appears to be a consequence of 
developing magnetic correlations. In contrast, the vortex-like Nernst signal 
does not persist nearly as far above $T_c$.

The above experiments on cuprates suggest that for a wide doping range the region 
above $T_c$ is a heterogeneous mixture of superconducting and magnetic correlations. 
A TF-$\mu$SR measurement is very sensitive to the spatially 
inhomogeneous magnetic response of such a system, due to the local-probe nature 
of the positive muon ($\mu^+$). The purpose of this review is to summarize the results
of recent {\it high} TF-$\mu$SR measurements on cuprates above $T_c$.
The experiments reveal a field-induced inhomogeneous line broadening that
persists far above $T_c$ and extends from the underdoped to heavily overdoped 
regimes. As will be explained in this review, the dominant source
of the line broadening is varied amongst the cuprates and is
dependent on a competition between superconducting and magnetic correlations.

\section{Transverse-field muon spin rotation (TF-$\mu$SR)}

In contrast to the closely related technique of nuclear magnetic resonance (NMR),
$\mu$SR does not rely on the Boltzmann population of spin states. Consequently,
application of the $\mu$SR method is not restricted by temperature, and an
external magnetic field is not required to polarize the muon spins. In fact
a powerful aspect of the $\mu$SR technique is that it can be used to probe
extremely weak internal magnetic fields in the absence of an overwhelming
external field. 
Zero-field (ZF) $\mu$SR has been applied extensively to cuprate
systems, and has played a pivotal role in determining where magnetism occurs 
in the temperature versus charge doping phase diagram of these materials
\cite{Weidinger:89,Niedermayer:98,Panagopoulos:02,Kiefl:89,Kanigel:02,Sanna:04,Miller:06}.
  
\begin{figure}
\centerline{\epsfxsize=5.0in\epsfbox{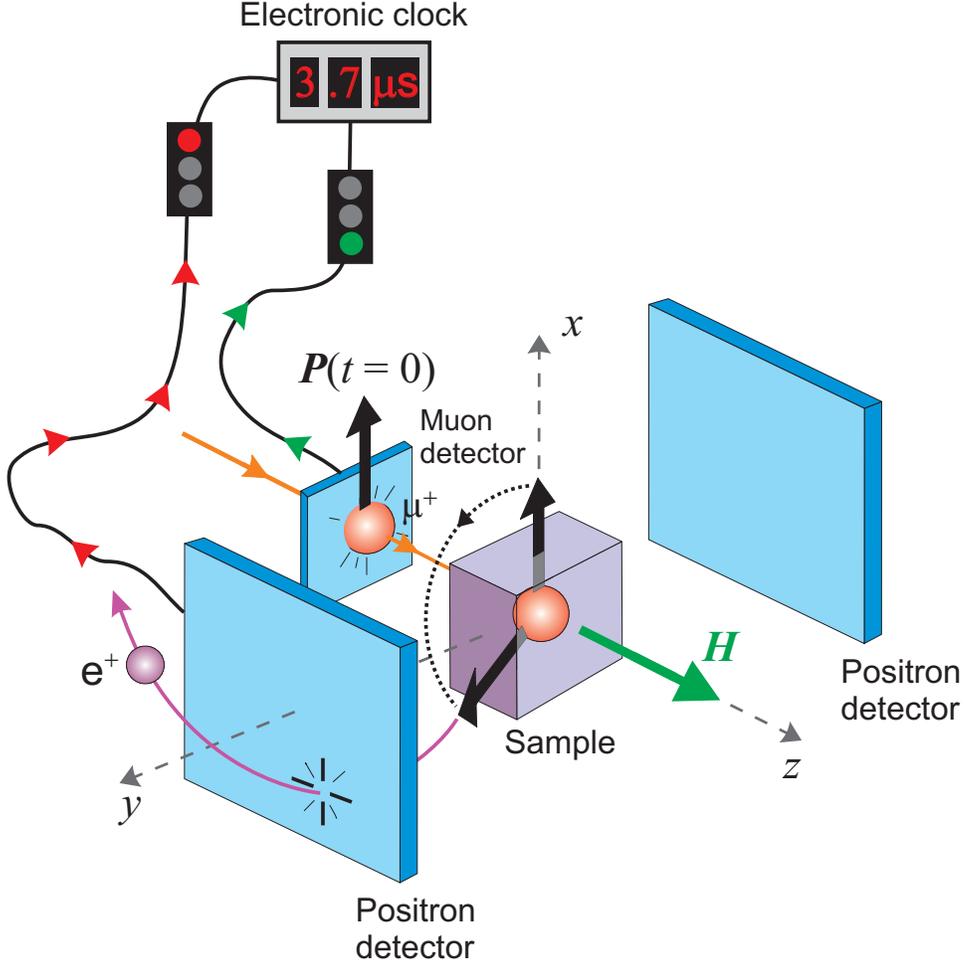}}
\caption{Basic layout of a high TF-$\mu$SR experiment. The beam of $\mu^+$
enters from the left with the initial muon spin polarization {\bf P}$(t = 0)$ 
perpendicular to the beam momentum. The external magnetic field {\bf H} 
is applied along the beam direction,
and hence perpendicular to {\bf P}$(t = 0)$. For cuprate samples comprised of plate-like
single crystals, the sample is usually mounted with the field 
parallel to the $c$-axis. This is so the larger $a$-$b$ area of the crystals exposes 
enough of the sample to the muon beam to achieve adequate count rates.
An incoming $\mu^+$ triggers the muon
detector, which starts a fast digital clock. The subsequent
detection of the decay positron stops the clock. For each such decay event,
the time interval is digitized and a count is added to the 
corresponding bin in a time histogram. Note the curved trajectory of the
decay positron that occurs in a high applied field.}
\label{figHighTFmuSR}
\end{figure}

The experiments of focus in this review are transverse-field (TF) $\mu$SR 
measurements, which locally probe the magnetic response of a material in the bulk.
In a TF-$\mu$SR experiment the direction of the initial muon spin polarizaton 
{\bf P}$(t = 0)$ is oriented perpendicular to the applied magnetic field,
as shown in figure~\ref{figHighTFmuSR}. The muon spin
precesses in the plane perpendicular to the direction of the effective 
local field {\bf B}$_{\mu}$, 
which in general is a vector sum of the applied field, dipolar fields of nearby nuclear and 
atomic moments, and the Fermi contact field of spin-polarized conduction electrons
at the muon site. The latter contribution is particularly important in metals, 
where the $\mu^+$ is screened by conduction electrons.  

\begin{figure}
\centerline{\epsfxsize=6.5in\epsfbox{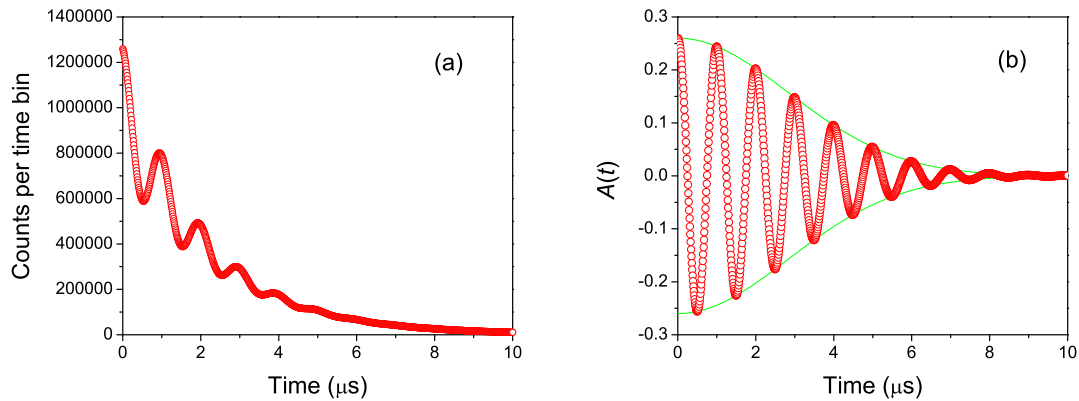}}
\caption{(a) Simulation of a raw time histogram $N(t)$ of a positron detector
in a TF-$\mu$SR experiment ({\it i.e.} equation~(\ref{eq:RawCounts})). 
The exponential decay is a 
consequence of the muon lifetime. The oscillation is due to the precession
of the muon spins in the field, which causes the muon spin polarization 
to sweep past the positron detector.
(b) Simulation of the corresponding asymmetry spectrum $A(t)$, as defined
in equation~(\ref{eq:asymmetry}). This example
shows a Gaussian loss of polarization with increasing time.}
\label{figRawCounts}
\end{figure}

In a $\mu$SR experiment one detects the 
positron emitted from the decay of the implanted positive muon 
($\mu^+ \! \rightarrow \! e^+ \! + \! \nu_e \! + \! \bar{\nu}_{\mu}$) using
scintillator detectors. The light produced when a muon or positron passes through
a thin plastic scintillator is transported by total internal reflection
through a plastic light guide to a photomultiplier tube. Fast electronics are
subsequently used to digitize, sort out, and store the electrical signals 
produced by the photomultiplier tubes.   
Each muon decay event is recorded as a function of time, resulting in a 
time-histogram of decay events given by

\begin{equation}
N_{\pm}(t)=N_{\pm}(0) e^{-t/\tau_\mu}[1 \pm a_0 G_{\rm TF}(t) \cos(\gamma_\mu B_{\mu} t + \phi)]    \, ,
\label{eq:RawCounts}
\end{equation}

\noindent where $N_+$ and $N_-$ are the count rates of the positron detectors positioned
on opposite sides of the sample,
$\tau_\mu \approx 2.2$~$\mu$s is the mean lifetime of the $\mu^+$,
$a_0 \leq 1/3$ is the initial asymmetry (dependent on the energy of the decay positrons and 
numerous experimental factors), $G_{\rm TF}(t)$ is the transverse muon spin depolarization function
(further discussed below), $\gamma_\mu \! = \! 0.0852$~$\mu$s$^{-1}$~G$^{-1}$ 
is the muon gyromagnetic ratio, $B_{\mu}$ is the magnitude of the local field, and
$\phi$ is the (phase) angle between the axis of the positron detector and the initial
muon spin polarizaton {\bf P}$(t = 0)$. An example of equation~(\ref{eq:RawCounts})
is shown in figure~\ref{figRawCounts}(a). 

For a surface muon beam, where the $\mu^+$
are generated via the decay of pions at rest in the laboratory frame of reference,
the initial muon spin polarization is nearly 100~\%. This very high degree of spin 
polarization is one of the reasons that $\mu$SR exceeds 
NMR in sensitivity to bulk magnetism. It is also important to emphasize that
in contrast to the majority of the NMR active nuclei in cuprates, 
the spin of the muon is 1/2, and hence the $\mu^+$ has no quadrupole moment.
Consequently, the $\mu^+$ is a ``pure'' magnetic probe.

The TF-$\mu$SR spectrum is the difference between the number of counts in 
the positron detectors positioned on opposite sides of the sample
divided by the sum of the counts in these detectors

\begin{eqnarray}
\frac{N_+ - N_-}{N_+ + N_-} = A(t) & = & a_0 G_{\rm TF}(t) \cos(\gamma_\mu B t + \phi) \nonumber \\
                                   & = & a_0 P(t) \, , 
\label{eq:asymmetry}
\end{eqnarray}

\noindent where $P(t)$ is the time evolution of the muon spin polarization. More
generally, $A(t)$ is referred to as the ``$\mu$SR asymmetry spectrum'', and is
analogous to an NMR free-induction decay signal. 
Figure~\ref{figRawCounts}(b) shows an asymmetry spectrum generated by 
equation~(\ref{eq:asymmetry}). The measured TF-$\mu$SR asymmetry spectrum has
error bars that increase with time due to the muon lifetime 
(see figure~\ref{figMacDougallPRB10}).

\subsection{Functional form of the depolarization}

The functional form of the depolarization function $G_{\rm TF}(t)$
is dependent on whether the effective field {\bf B}$_{\mu}$ experienced 
by the muon is static or fluctuating in time: \\ 

\noindent {\it Static fields} --- If the muons experience an inhomogeneous distribution of dipolar 
fields from static nuclear or electronic moments, or the inhomogeneous field distribution
created by static vortices in the superconducting state,
muons stopping at different locations
experience different local magnetic fields. Consequently, the muons 
precess with different frequencies,  
given by $\nu_{\mu} \! = \! \gamma_{\mu} B_{\mu}/ 2 \pi$. 
Over time the muon spins precessing
at different frequencies dephase, causing a loss of polarization.

The host nuclei usually form a dense system of randomly oriented moments that are 
static on the $\mu$SR time scale and create a Gaussian distribution of 
dipolar fields at the muon site. This results in a Gaussian     
depolarization function of the form

\begin{equation} 
G_{\rm TF}(t) = \exp(-\sigma^2 t^2/2) \, , 
\end{equation}

\noindent where $\sigma^2 \! = \! \gamma^2_{\mu} \langle B^2_{\mu} \rangle$, 
with $\langle B^2_{\mu} \rangle$
being the width of the Gaussian field distribution along the direction of the applied field.
An asymmetry spectrum with a Gaussian loss of 
polarization is shown in figure~\ref{figRawCounts}(b). 
Barring a change in crystal structure or the onset of thermally-induced muon hopping, 
the depolarization rate $\sigma$ associated with the nuclear moments does not depend 
on temperature. On the other hand, $\sigma$ can depend on both the direction and strength
of the applied magnetic field. In the latter situation, 
the nuclear moments may rapidly precess about the applied field, resulting in a 
reduction of $\sigma$.

A dense system of randomly frozen electronic moments will also lead to a Gaussian 
depolarization function. On the other hand, a dilute system of random electronic
moments is decribed by an exponential depolarization function that corresponds
to a Lorentzian distribution of internal fields \cite{Walker:74}

\begin{equation} 
G_{\rm TF}(t) = \exp(-\Lambda t) \, . 
\end{equation}

\noindent {\it Fluctuating fields} --- The Gaussian depolarization observed for 
a dense system of static moments is modified by fluctuations. Fast fluctuations
of the internal fields results in an exponential loss of the muon spin polarization,
such that

\begin{equation}
G_{\rm TF} = \exp(-\sigma^2 \tau t) = \exp(-\Lambda t) = \exp(-t/T_2) \, ,
\end{equation}

\noindent where $1/\tau$ is the fluctuation rate of {\bf B}$_{\mu}$. The exponential
relaxation rate $\Lambda \! = \! \sigma^2 \tau \! \equiv \! 1/T_2$ becomes
smaller as the fluctuation rate rises. 
This is referred to as ``motional narrowing'' --- a term borrowed from NMR
where one generally works in frequency space. It means that the width of the 
Fourier transform of the TF-$\mu$SR signal is reduced as fluctuations of the
local field increase. 

It is not always obvious whether the exponential loss of polarization is caused by
fast fluctuating internal fields or a dilute system of static moments. In these instances
a longitudinal field (LF) experiment can be performed by applying a field parallel
to the direction of the initial muon spin polarization. 
If the internal fields are static, a longitudinal field of 
greater magnitude will decouple the muon spin from the
internal fields and no relaxation of the LF-$\mu$SR signal will be observed.
Alternatively, fast fluctuating internal fields cause a ($T_1$) relaxation of 
the muon spin polarization parallel to the external field, by inducing transitions
between the ``spin up'' and ``spin down'' Zeeman-energy eigenstates of the muon.
In this case a much higher external field is needed to decouple 
the muon spin from the internal fields.
\\ 

\noindent {\it Disorder effects} --- In a system with macroscopic phase segregation,
the muon will experience different local magnetic fields in the spatially-separated 
regions. In this case the TF-$\mu$SR signal will be comprised of well-resolved muon 
precession frequencies. Moreover, the amplitudes of the different frequency components 
of the TF-$\mu$SR signal, provide a direct measure of the volume fraction of each phase.
Phase segregation may occur on a smaller length scale with disorder causing
a spread in muon precession frequencies. 
Of particular interest in this review is the muon depolarization rate
resulting from an inhomogeneous distribution of the magnetic susceptibility.
A inhomogeneous spread $\delta \chi$ of static or time-averaged
local susceptibilities $\chi$ produces 
a distribution of $\mu^+$ Knight shifts $\delta K = \delta(B - H)/H$, and 
hence a broadening of the 
TF-$\mu$SR line width \cite{MacLaughlin:96}. It follows that

\begin{equation}
\delta(B - H) = \delta K \, H \propto \delta \chi \, H \, .
\label{Knight}
\end{equation}   

\noindent In this situation the different components of the TF-$\mu$SR signal are numerous
and cannot be isolated, but the depolarization rate of the signal is 
a direct measure of the width of the inhomogeneous field distribution.

\subsection{Technical limitations of TF-$\mu$SR at high magnetic field}

At present the only functioning high TF-$\mu$SR spectrometer in the world
is {\it HiTime}, located at TRIUMF (Vancouver, Canada). The {\it HiTime}
spectrometer utilizes a 7 T superconducting magnet, with detectors optimized
for high transverse field. Several technical constraints have to be
overcome to perform a TF-$\mu$SR experiment at such high field: \\  

\noindent (i) Surface muon beams created from pion decay at rest in the laboratory frame, 
result in an initial muon spin polarization that is anti-parallel to the linear 
momentum of the beam. A magnetic field applied perpendicular to the momentum
produces a Lorentz force, which deflects the muon beam. The degree of deflection
becomes problematic for fields in excess of $0.01$~T.
To circumvent this problem the muon spin can be rotated perpendicular to the
momentum, in which case the external magnetic field may be directed parallel
to the incoming beam. This can be done only on muon beam lines equipped with 
a ``spin rotator'', which is a device consisting of crossed electric 
and magnetic fields. \\
     
\begin{figure}
\centerline{\epsfxsize=4.5in\epsfbox{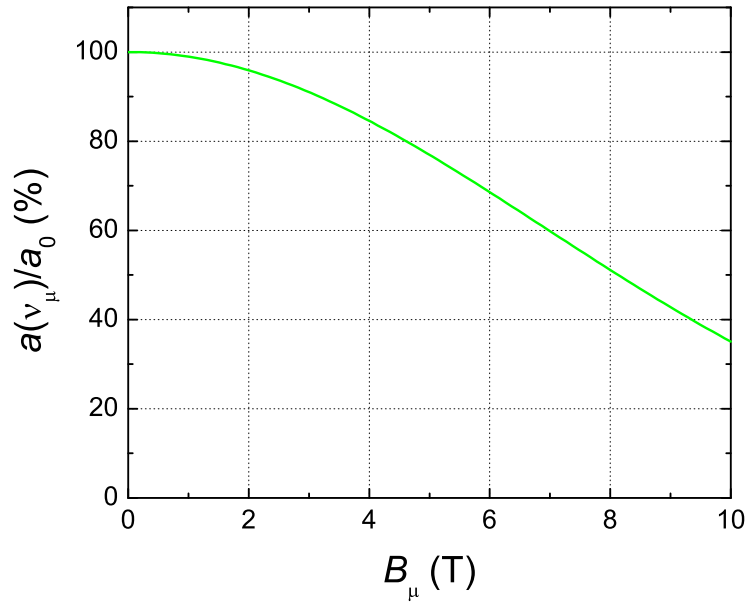}}
\caption{The amplitude of the TF-$\mu$SR signal as a function of field
for a timing resolution of 170~ps.}
\label{figTimeResol}
\end{figure}

\noindent (ii) At high fields the muon-spin precession period 
$T_{\mu} = 1/\nu_{\mu} \! = \! 2 \pi /\gamma_{\mu} B_{\mu}$ is reduced,
so that a smaller bin width must be used in the time histogram
to generate the raw time spectrum $N(t)$. However, the timing resolution 
of both the electronic circuitry and the detectors limits how small the bin width
can be, and this in turn limits the maximum field measurable by TF-$\mu$SR.
With increasing magnetic field, the initial amplitude of the TF-$\mu$SR 
signal decays according to \cite{Holzschuh:83}

\begin{equation}
\frac{a(\nu_{\mu})}{a_0} = 
\exp \left[ \frac{-(\pi 2.355 \nu \Delta t)^2}{4 \ln 2} \right] \, ,
\end{equation}

\noindent where $a(\nu_\mu)$ is the amplitude of the TF-$\mu$SR signal of 
precession frequency $\nu_\mu$, and $\Delta t$ is the timing resolution.
Figure~\ref{figTimeResol} shows the field dependence of the signal amplitude
of the 7~T {\it Hitime} spectrometer at TRIUMF, which
has a time resolution of about 170~ps. At 7 T the signal amplitude
is reduced to 60~\% of the low-field value. \\

\noindent (iii) The incoming muons precess in the transverse field prior to implantation.
If the field is uniform, all of the muon spins precess in phase, and the initial
muon-spin polarization is preserved. However, if the field is non-uniform 
({\it i.e.} there are field components perpendicular to the 
muon beam momentum), the initial
polarization is reduced by dephasing of the muon spins. Thus, high TF-$\mu$SR
experiments require superconducting 
magnets with good field homogeneity to be used in combination 
with muon beams of small cross section. Unfortunately, to preserve good timing resolution
the light guides that connect the plastic scintillators to the photomultiplier tubes 
outside the magnet bore must be kept short to minimize attenuation and
broadening of the light pulses. This means that 
the cylindrical bore of the magnet must be kept short. But a short bore limits
the field homogeneity of the magnet. In principle this limitation can be
overcome by installing shim coils. \\

\noindent (iv) At high magnetic field the decay positrons undergo cyclotron motion
about the applied field directed along the beam axis. The radius of the
helical trajectories of the positrons is inversely proportional to the field.
Hence at high field the positron detectors must be positioned closer 
to the sample. For surface
muons that have a momentum of about 28~MeV/c, the cyclotron radius of the
positrons at 7~T is only 1.3~cm. For this reason the {\it HiTime} spectometer utilizes
a unique compact arrangement of muon and positron counters
contained within a helium-gas flow cryostat. \\   

As a result of the above considerations, high TF-$\mu$SR instruments are
not widely available. With such limited access, high 
TF-$\mu$SR experiments on cuprates (using {\it HiTime} at TRIUMF) 
have primarily focussed on the superconducting state. 
But in recent years
such studies have shifted attention to examining the effects of
high magnetic field on the normal state above $T_c$, and
this has led to some surprising results. 
        
\section{Field-induced/enhanced magnetic order below $T_c$}

Within the underdoped regime of cuprates, an applied magnetic field 
has been shown to induce static magnetic order and enhance magnetic order 
that is already present at $H \! = \!0$.
Demler \etal \cite{Demler:01} introduced a phenomenological quantum theory
that describes this field-induced/enhanced effect as 
a competition between superconductivity and 
magnetic order. The original model assumes a two-dimensional (2D) system
that undergoes a field-induced quantum phase transition (QPT) from a
superconductor to a state of coexisting superconducting and spin-density wave
(SDW) orders. In 2D, long range SDW order does not survive above $T \! = \! 0$,
and hence the model strictly applies to the situation at low temperatures.
The dependence of the QPT on charge-carrier concentration in the theory
of Demler \etal \cite{Demler:01} is inline with low-temperature
neutron scattering experiments on underdoped cuprates that show field-induced 
and/or field-enhanced SDW order in hole-doped
\cite{Katano:00,Lake:02,Khaykovich:05,Chang:08,Haug:09} and electron-doped
\cite{Matsuura:03,Kang:05} systems.
Recently, Moon and Sachdev \cite{Moon:09} have derived a microscopic theory of
competing $d$-wave superconductivity and SDW order that
reproduces the key features of the phase diagram of the phenomenological
theory of Demler \etal \cite{Demler:01}. 

Field-induced/enhanced competing SDW order is also detectable by TF-$\mu$SR, and has
been observed in both hole-doped \cite{Savici:05,Sonier:09} and electron-doped 
\cite{Sonier:03} systems. However, as explained above, 
a shortcoming of TF-$\mu$SR is the maximum external field of 7 T that can be applied.     
This is particularly an issue in the YBa$_2$Cu$_3$O$_y$ family where superconductivity
is more robust, and consequently higher fields are required to expose or enhance
the SDW order. For example, neutron scattering experiments by Stock 
\etal \cite{Stock:09} on YBa$_2$Cu$_3$O$_{6.33}$ and 
YBa$_2$Cu$_3$O$_{6.35}$ show that a 6 T field directed along the $c$-axis
is insufficient to enhance static magnetic order via the suppression of the 
superconducting order parameter. Field enhanced SDW order in YBa$_2$Cu$_3$O$_{6.45}$
was recently reported by Haug \etal \cite{Haug:09}, but an external
field having a $c$-axis component with a magnitude of 12.5~T was used.
While neutron scattering is better suited for investigations of long-range SDW order, 
as a local probe TF-$\mu$SR measurements can furnish information on
short-range magnetic order, disordered magnetism, and the magnetic volume fraction.
      
\begin{figure}
\centerline{\epsfxsize=4.5in\epsfbox{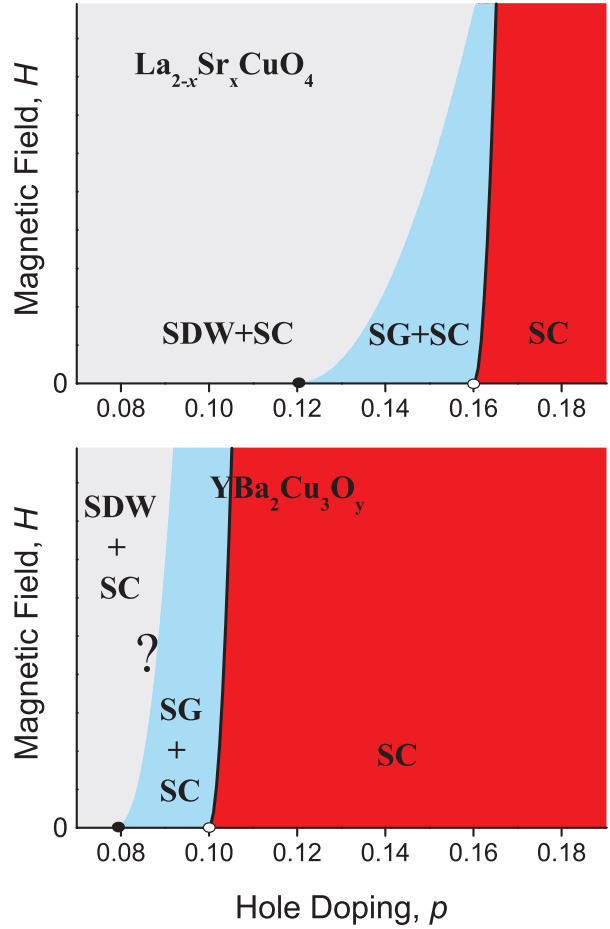}}
\caption{$H$-versus-$p$ phase diagrams for La$_{2-x}$Sr$_x$CuO$_4$
and YBa$_2$Cu$_3$O$_y$ at $T = 0$. The QPT transition to a state 
with spin-glass-like
magnetism coexisting with superconductivity (SG+SC) is deduced from
low TF-$\mu$SR measurements \cite{Sonier:07}. For La$_{2-x}$Sr$_x$CuO$_4$, 
neutron scattering measurements \cite{Khaykovich:05,Chang:08}
indicate that there is a crossover to a state where long-range 
static SDW order coexists with superconductivity (SDW+SC). 
This is also observed in YBa$_2$Cu$_3$O$_{6.45}$ \cite{Haug:09},
but the hole-doping dependence of this crossover has not yet been determined
for YBa$_2$Cu$_3$O$_y$.  
According to Kivelson \etal \cite{Kivelson:02}, the phase transition
ends at an ``avoided'' quantum critical point (open circle).
The ``true'' quantum critical point is at a lower hole-doping 
concentration (solid black circle),
which apparently coincides with the critical doping at which static magnetism
vanishes at $H \! = \! 0$ \cite{SonierRPP:07}.}
\label{PhaseDiagrams}
\end{figure}

\subsection{Detection of field-induced spin-glass like magnetism by $\mu$SR}

In the 2D theory of Demler \etal \cite{Demler:01}, field-induced SDW order 
occurs at $T \! = \! 0$ in and around the vortex cores 
where superconductivity is suppressed. Long-range SDW order requires significant 
overlap of neighbouring vortices. The density of vortices and hence the degree 
of overlap increases with field.   
But the applicability of a 2D theory to the cuprates is limited by
the significant coupling between the CuO$_2$ layers in the real materials.  
Experiments on La$_{1.90}$Sr$_{0.10}$CuO$_4$ show that the vortices 
\cite{Divakar:04} and the field-enhanced SDW order \cite{Lake:05}  
are in fact three-dimensional (3D). Kivelson \etal \cite{Kivelson:02} 
have extended the phenomenological theory of Demler \etal to the case 
of 3D vortices and shown that the QPT is actually to a phase in 
which a spatially inhomogeneous competing order coexists with superconductivity.
At the QPT static magnetism is expected to develop about weakly interacting 
vortices, but there is no long-range static magnetic order.
A crossover to long-range SDW order occurs at higher field and/or 
lower charge carrier concentration. The situation is summarized in 
figure \ref{PhaseDiagrams}.

\begin{figure}
\centerline{\epsfxsize=6.5in\epsfbox{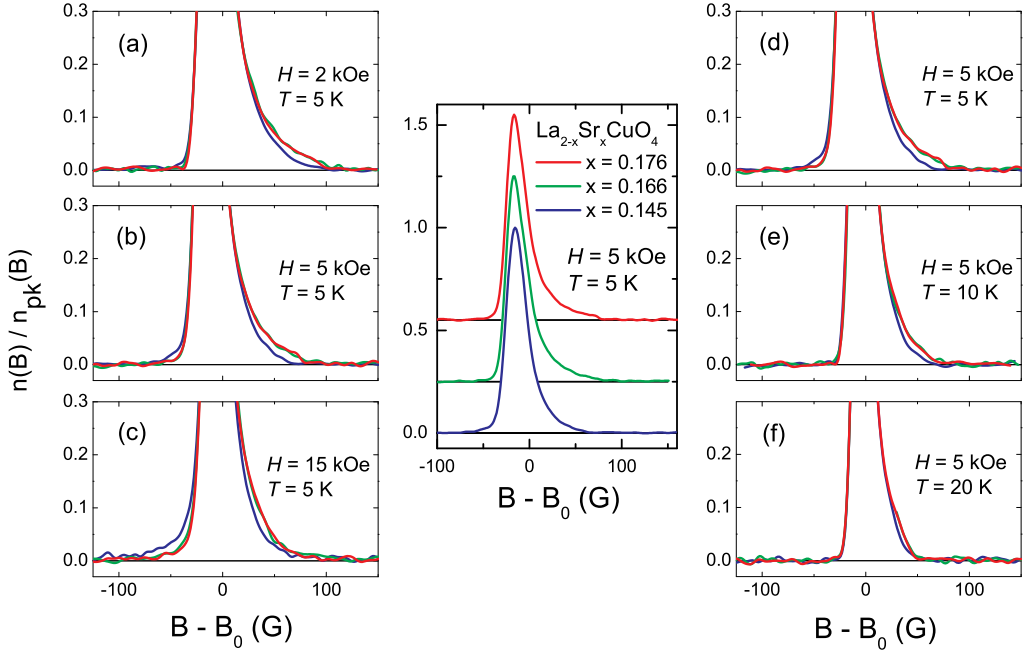}}
\caption{Doping, temperature and magnetic field dependences 
of the $\mu$SR line shapes for La$_{2-x}$Sr$_x$CuO$_4$ with 
$x \! = \! 0.145$, 0.166 and 0.176 (from Ref~\cite{Sonier:07}).
The center panel shows full $\mu$SR line shapes at $H \! = \! 5$~kOe
and $T \! = \! 5$~K. Panels (a)-(c) show the field dependence of the `tail' regions 
of the normalized $\mu$SR line shapes. Panels (d)-(f) show the temperature 
dependence of the `tail' regions of the normalized $\mu$SR line shapes.}
\label{figSonierPRB07}
\end{figure}

The extended 3D model of Kivelson \etal is supported by 
low-field TF-$\mu$SR measurements, which detect static
spin-glass-like (SG) magnetism in and around the vortex cores of
La$_{2-x}$Sr$_x$CuO$_4$ and YBa$_2$Cu$_3$O$_y$ \cite{Sonier:07}.
The static magnetism in the vortex-core region causes a suppression 
of the high-field `tail' of the internal magnetic field distribution $n(B)$.
If the distribution of the dipolar fields of the electronic moments
is sufficiently broad, this is accompanied by the appearance of a low-field tail. 
An example of these field-induced effects are shown 
in figure~\ref{figSonierPRB07} for La$_{2-x}$Sr$_x$CuO$_4$ at doping levels where
static electronic moments are absent at $H \! = \! 0$. 
At $H \! < \! 1.5$~T the samples with 
$x \! = \! 0.166$ and $x \! = \! 0.176$ are in the pure SC phase 
depicted in the top panel of figure~\ref{PhaseDiagrams}. 
The $\mu$SR line shape, which is a
Fourier transform of $P(t)$ and a close representation of $n(B)$, is 
nearly identical for these two samples. Conversely, the
$\mu$SR line shape at $x \! = \! 0.145$ shows modifications of the high-field 
and low-field tails expected from the emergence of SG magnetism in
the vortex-core region, which places this sample in the SG+SC
phase shown in figure~\ref{PhaseDiagrams}. With increasing temperature the 
$\mu$SR line shape at $x \! = \! 0.145$ approaches that for the higher
doped samples (see figures \ref{figSonierPRB07}(d)-(f)), and just below $T_c$ the 
line shapes at the different dopings are nearly indistinguishable.
This demonstrates the destruction of the static magnetism at
$x \! = \! 0.145$ by thermal fluctuations. Consistent with the 3D phenomenological
theory of Kivelson {\it et al}, the SG+SC phase deduced by TF-$\mu$SR
occurs above the hole doping concentration 
at which static magnetism is observed at $H \! = \! 0$ \cite{SonierRPP:07}.

\subsection{Detection of field-induced static magnetic order by $\mu$SR}

As mentioned, long-range SDW order occurs in the superconducting phase 
(SDW+SC in figure~\ref{PhaseDiagrams}) when 
there is a strong overlap of the vortices. A signature of strong overlap 
in a $d_{x^2-y^2}$-wave superconductor is the formation of a square vortex
lattice \cite{Ichioka:99}. A square vortex lattice is observed in La$_{2-x}$Sr$_x$CuO$_4$
\cite{Gilardi:02} at much lower fields than in YBa$_2$Cu$_3$O$_y$ \cite{Brown:04}.
This is consistent with the higher field needed to induce or enhance long-range 
SDW order in YBa$_2$Cu$_3$O$_y$ \cite{Haug:09}. TF-$\mu$SR measurements of 
La$_{2-x}$Sr$_x$CuO$_4$ in the SDW+SC regime show a substantial 
{\it fast} relaxing component
associated with the SDW order \cite{Chang:08,Savici:05,Sonier:09,Sonier:08}.
The fast relaxation indicates a varying degree of SDW order due to the spatial
variation of the superconducting order parameter in the vortex state. 
The TF-$\mu$SR measurements on La$_{2-x}$Sr$_x$CuO$_4$ show that the onset of
SDW order occurs at a temperature well below $T_c$, whereas the neutron scattering
experiments indicate that the SDW order sets in close to
$T_c$. This difference is apparently a consequence of the different time 
scales of the two techniques. The theories of competing magnetic and superconducting
orders do not explicitly treat the situation at finite temperature. However,
thermal fluctuations are expected to cause fluctuations of the SDW order, so that
static long-range SDW order exists only below $T_c$ \cite{Moon:09}.   

\section{Field-induced inhomogeneous line broadening above $T_c$}

The normal state of La-based cuprates above $T_c$ was first studied with high
TF-$\mu$SR by Savici \etal \cite{Savici:05}. 
The TF-$\mu$SR signals for 
La$_{1.88}$Sr$_{0.12}$CuO$_4$, La$_{1.875}$Ba$_{0.125}$CuO$_4$ and 
La$_{1.75}$Eu$_{0.1}$Sr$_{0.15}$CuO$_4$ exhibited a field-enhanced exponential 
relaxation rate that extended far above $T_c$. These samples all showed
evidence of static electronic moments in ZF-$\mu$SR measurements.   
On the other hand, no field-induced relaxation of the TF-$\mu$SR signal
was observed above $T_c$ in 
optimally-doped (Bi, Pb)$_2$Sr$_2$CaCu$_2$O$_8$, overdoped
La$_{1.81}$Sr$_{0.19}$CuO$_4$ or YBa$_2$(Cu$_{2.979}$Zn$_{0.021}$)O$_7$ samples that
do not exhibit static magnetism at $H \! = \! 0$. Hence the field-induced
relaxation rate could be attributed to field-enhanced magnetism.  
However, subsequent high TF-$\mu$SR studies have shown that this 
field-induced effect above $T_c$ also takes place in La$_{2-x}$Sr$_x$CuO$_4$
samples of higher Sr (hole doping) concentration \cite{Sonier:08,MacDougall:06,MacDougall:10}, 
and underdoped samples of YBa$_2$Cu$_3$O$_y$ \cite{Sonier:08}, in which the 
electronic moments fluctuate at a rate outside the $\mu$SR time window at $H \! = \! 0$. 
    
In the SDW+SC regime of La$_{2-x}$Sr$_x$CuO$_4$, 
the TF-$\mu$SR depolarization function is observed to be of 
the form \cite{Savici:05,Sonier:09,Sonier:08} 

\begin{equation}
G_{\rm TF}(t) = [(1-f)e^{-\Lambda t} + fe^{-\lambda t}]e^{-\sigma^2 t^2} \, .
\label{eq:Gxt}  
\end{equation}

\noindent The Gaussian function accounts for the random nuclear dipole fields 
and is temperature independent. The two-component exponential function in square
brackets implies that there are two spatially separated sources of
the depolarization.
A fraction $f$ of the muons stopping in the sample sense the
$\lambda$ component, which is associated with SDW correlations stabilized in 
and around the vortex cores. The near exponential form of this 
depolarization is apparently a consequence of the spatial separation
of the vortex cores containing short-range SDW order.
In the SG+SC phase depicted in figure~\ref{PhaseDiagrams}, the magnetic volume fraction
associated with the spin-glass like magnetism in the vortex core region is
too small to be resolved as a distinct component of the depolarization function.   
Thus far the $\lambda$ component has been observed
in underdoped La$_{1.88}$Sr$_{0.12}$CuO$_4$ \cite{Savici:05} and
La$_{1.855}$Sr$_{0.145}$CuO$_4$ \cite{Sonier:09}, where the
field required to induce static SDW order is attainable with high 
TF-$\mu$SR.
On the other hand, the exponential depolarization rate $\Lambda$ of 
equation~(\ref{eq:Gxt}) is observed in both the underdoped and overdoped regimes
of La$_{2-x}$Sr$_x$CuO$_4$. It has even been
observed in heavily overdoped non-superconducting La$_{2-x}$Sr$_x$CuO$_4$ 
\cite{MacDougall:06,MacDougall:10,Sonier:10}. An analogous field-induced
exponential decay of the muon spin polarization has also been detected in
superconducting YBa$_2$Cu$_3$O$_y$ at $H \! = \! 7$~T \cite{Sonier:08}.

\begin{figure}
\centerline{\epsfxsize=4.5in\epsfbox{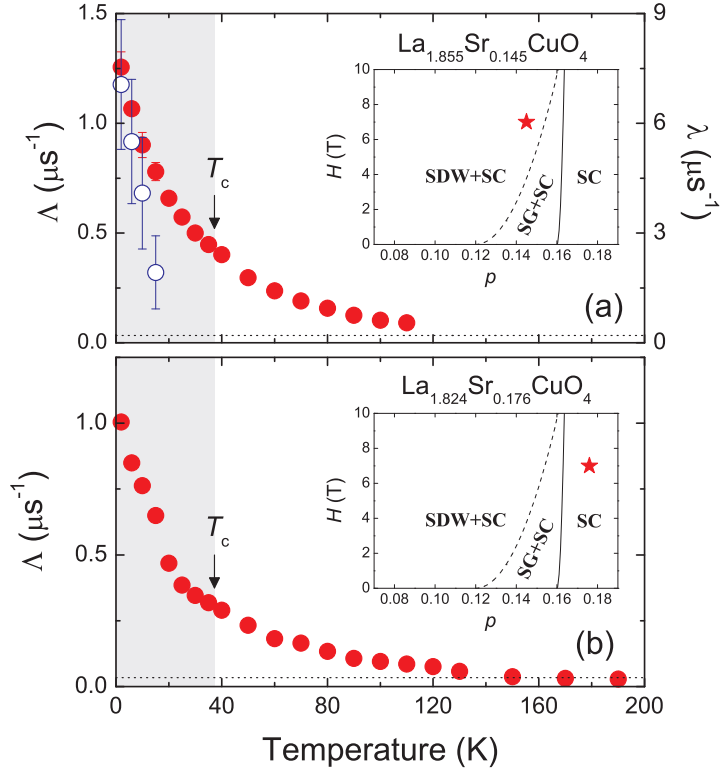}}
\caption{Temperature dependence of the exponential depolarization rates
$\lambda$ (open circles) and $\Lambda$ (solid circles) in
(a) La$_{1.855}$Sr$_{0.145}$CuO$_4$, and (b) La$_{1.824}$Sr$_{0.176}$CuO$_4$  
at $H \! = \! 7$~T.
The horizontal dotted lines indicate the contribution to $\Lambda$
from the inhomogeneity of the external field.
The bulk superconducting transition temperature at zero applied
field is denoted by $T_c$.
The inset for each panel shows the location (indicated by a star) 
in the $H$-versus-$p$ phase diagram at $T \! \ll \! T_c$. 
Note the solid curves denoting the SC-to-SG+SC phase transition and the 
dashed curves denoting the SG+SC-to-SDW+SC crossover are approximate.}
\label{fig6_TR}
\end{figure}

As an example, figure~\ref{fig6_TR} shows the temperature dependence of 
the exponential depolarization rates in underdoped and overdoped 
La$_{2-x}$Sr$_x$CuO$_4$ at $H \! = \! 7$~T.
The {\it faster} relaxing $\lambda$ component is present only in the underdoped 
($x \! = \! 0.145$) sample, and does not persist above $T_c$. This indicates 
that static SDW order exists at this
doping and field only in the superconducting state. At both dopings the {\it slower}
relaxing $\Lambda$ component persists well above $T_c$. Note that in some studies
a different notation is used, and the depolarization rate extending far 
above $T_c$
is denoted by $\lambda$ (rather than $\Lambda$). Also, at low temperatures,
especially in the vortex state, the depolarization at high field is only
approximately of exponential form.   
 
\subsection{Lack of dynamics}

The TF-$\mu$SR experiments measure the combined effects of static 
and dynamic (time-varying) local fields on the internal magnetic field distribution 
--- meaning that both static-field 
inhomogeneity and fluctuating local fields may contribute to the
decay rate $\Lambda$.
Because $\mu$SR can detect relaxation times below the lower limit 
of the NMR time window, $\mu$SR is better suited for resolving slow fluctuations
of the local field. However, it is not always easy to distinguish depolarization of the 
TF-$\mu$SR signal due to an inhomogeneous static field distribution from
relaxation caused by fluctuating internal fields.  

Savici \etal \cite{Savici:05} and MacDougall \etal \cite{MacDougall:06} have
investigated the dynamics of the field-induced depolarization in 
La$_{2-x}$Sr$_x$CuO$_4$ above $T_c$ by LF-$\mu$SR. 
In a LF-$\mu$SR experiment, the external field
is applied parallel to the initial direction of the muon spin polarization.
When the applied longitudinal field exceeds the strength of the 
internal static fields, any spatially inhomogeneous distribution of
static field has no effect on the time evolution of the muon spin polarization.
In this situation, the LF-$\mu$SR signal is still susceptible to relaxation
caused by rapidly fluctuating internal fields. When the fluctuation rate $\nu$ of the
internal fields $B(t)$ is such that $\nu/\Delta \gg 1$, where $\Delta/\gamma_\mu$ is 
the root mean square (rms) of $B_i(t)$ ($i = x, y, z$), the longitudinal
relaxation function has an exponential decay

\begin{equation}
G_{\rm LF}(t) = e^{-t/T_1} \, ,
\end{equation}            

\noindent where $1/T_1$ is the {\it longitudinal} or $T_1$ relaxation rate given by

\begin{equation}
\frac{1}{T_1} = \frac{2 \Delta^2/ \nu}{1 + (\gamma_\mu H_z/ \nu)^2} \, .
\end{equation}

However, measurements on underdoped \cite{Savici:05} and overdoped
\cite{MacDougall:06} La$_{2-x}$Sr$_x$CuO$_4$ show no relaxation of the LF-$\mu$SR 
signal, even at temperatures far above $T_c$. Taking into
account the accuracy of their experiment, Savici \etal \cite{Savici:05} set 
a conservative upper limit of $1/T_1 < 0.05$~$\mu$s$^{-1}$. 
Combined with the observation that $\Lambda \gg 1/T_1$, they have ruled out fields 
fluctuating at a rate greater than $10^9$~Hz.
Hence the field-induced relaxation rate $\Lambda$ in La$_{2-x}$Sr$_x$CuO$_4$ 
is interpreted as a broadening of an inhomogeneous static
field distribution. Likewise, LF-$\mu$SR measurements on YBa$_2$Cu$_3$O$_{6.57}$
single crystals in fields up to $H \! = \! 5.5$~T show no evidence for
dynamic relaxation \cite{Sonier:98}.         

\subsection{Relation to the vortex-like Nernst signal} 

Ong \etal \cite{Xu:00,Wang:01,Wang:06} interpret the large Nernst signal 
they observe above $T_c$ as being the response of a 2D vortex liquid. 
Like the relaxation rate $\Lambda$, the vortex-like Nernst signal decays 
continuously across $T_c$ and gradually decreases with increasing temperature. 
However, while TF-$\mu$SR can be used to detect the onset of a vortex solid-to-liquid 
transition, it is generally insensitive to variations of the vortex motion
within the liquid phase itself. 
This is a consequence of the microsecond $\mu$SR time scale.
Upon entering the vortex liquid state, the local magnetic field distribution 
measured by TF-$\mu$SR is severely narrowed by the rapid motion of the vortices 
\cite{Brandt:91}. Once this occurs, further increases in vortex mobility have 
no appreciable effect on the measured field distribution --- as amply demonstrated 
by numerous TF-$\mu$SR studies of vortex-liquid phases below $T_c$ \cite{Sonier:00}.

\begin{figure}
\centerline{\epsfxsize=5in\epsfbox{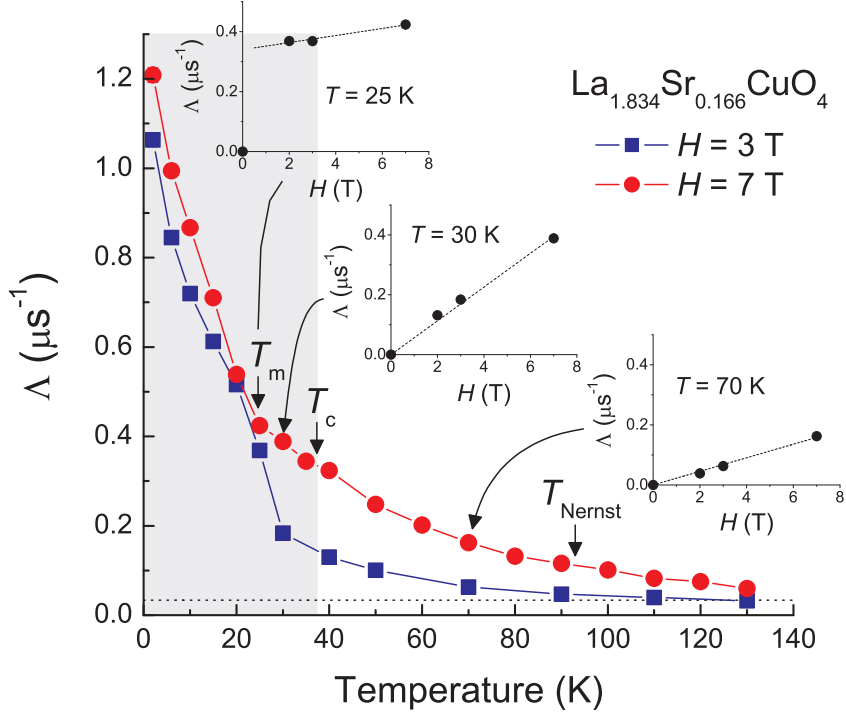}}
\caption{Temperature dependence of the exponential depolarization rate
$\Lambda$ in La$_{1.834}$Sr$_{0.166}$CuO$_4$ 
at $H \! = \! 3$ and 7~T \cite{Sonier:08}. 
Above $T_m$, $\Lambda$ is proportional to $H$.
The onset of the Nernst signal is denoted by 
$T_{\rm Nernst}$ \cite{Wang:01}, and $T_c = 37.3$~K corresponds to the
bulk $\Lambda$ below $T_c$superconducting transition temperature at $H \! = \! 0$.
The horizontal dotted line indicates the contribution of the 
inhomogeneity of the external field to $\Lambda$.}
\label{fig7_TR}
\end{figure}

In the high TF-$\mu$SR experiments 
\cite{Savici:05,Sonier:08,MacDougall:06,MacDougall:10,Sonier:10}, 
a vortex liquid is definitively ruled out as the source of $\Lambda$ by the measured
field dependence. Contrary to a vortex liquid, the width of the 
local field distribution (which is proportional to $\Lambda$) 
above and immediately below $T_c$ is observed to increase with field. 
At the maximum field of $H \! = \! 7$~T, $\Lambda$ persists above the Nernst 
region (see figure~\ref{fig7_TR}). Here it must be pointed out that 
the Nernst signal above $T_c$ consists of a strongly field-dependent
contribution from superconducting fluctuations, and a field-independent
contribution from normal-state quasiparticles \cite{Behnia:09}. 
Recently Cyr-Choini\'{e}re \etal \cite{Cyr:09} have shown that the contribution
due to superconducting fluctuations does
not extend nearly as far above $T_c$ as was assumed from the original
experiments by Ong \etal \cite{Xu:00,Wang:01,Wang:06}. They find that
in La$_{1.875}$Sr$_{0.125}$CuO$_4$, and also in La$_{2-x}$Sr$_x$CuO$_4$ doped 
with Eu or Nd, the Nernst signal that persists at temperatures well
above the regime of superconducting fluctuations is caused 
by ``stripe-like'' charge and spin order.
Similar conclusions have been reached in a Nernst study of YBa$_2$Cu$_3$O$_y$
by Daou \etal \cite{Daou:09}, where it has also been shown that the onset
of the Nernst signal coincides with the pseudogap temperature $T^*$.
 
Despite these distinctions, there does appear to be some connection between 
the TF-$\mu$SR and Nernst experiments. The vortex-like Nernst signal extends
into the vortex liquid state below $T_c$. Likewise,
$\Lambda$ first develops an appreciable field dependence below $T_c$. Figure~\ref{fig7_TR}
displays the temperature and magnetic field dependences of $\Lambda$
for overdoped La$_{1.834}$Sr$_{0.166}$CuO$_4$. Below a temperature $T_m$, 
$\Lambda$ has a weak field dependence (for fields above the lower critical field
$H_{c1}$) that is indicative of static disorder in a vortex glass phase \cite{Divakar:04}.
At temperatures above $T_m$, $\Lambda$ is observed to be proportional to $H$
\cite{Savici:05,Sonier:08,MacDougall:06,MacDougall:10}, 
and this behaviour persists until $\Lambda$ saturates well above $T_c$.
It is difficult to tell from existing measurements whether the 
$\Lambda \! \propto \! H$ behaviour begins precisely at the onset of the vortex 
liquid phase below $T_c$. However, there is certainly a temperature range below $T_c$ where
$\Lambda \propto H$ behaviour and a vortex-like Nernst signal are simultaneously observed.

\section{Variations in the hole-doping dependence}

Numerous TF-$\mu$SR studies of cuprates have been devoted to measurements of the
hole-doping dependence of the depolarization rate below $T_c$ at low field 
({\it i.e.} in the vortex state).
From such studies a universal linear relation between $T_c$ and the muon spin
depolarization rate was obtained in the underdoped regime --- the so-called 
``Uemura plot'' \cite{Uemura:89}. The depolarization rate in the vortex state is
dominated by the spatial field inhomogeneity created by the arrangement of vortices, 
which in turn
is proportional to $\lambda_{ab}^{-2}$, where $\lambda_{ab}$ is the in-plane 
magnetic penetration depth. Because $\lambda_{ab}^{-2} \! \propto \! \rho_s$, where
$\rho_s$ is the density of superconducting carriers, the Uemura plot is considered
to be a scaling relation between $T_c$ and $\rho_s$. 

Not much has been learned from low TF-$\mu$SR measurements above $T_c$, where the
muon spin depolarization rate is dominated by the nuclear dipole fields. However,
the most striking and revealing aspect of recent high TF-$\mu$SR measurements on 
hole-doped cuprates is the doping dependence of $\Lambda$ above $T_c$. This turns out
to be very different in the La$_{2-x}$Sr$_x$CuO$_4$ and YBa$_2$Cu$_3$O$_y$ systems.  

\subsection{La$_{2-x}$Sr$_x$CuO$_4$: Evidence for diverse forms of magnetism}

\begin{figure}
\centerline{\epsfxsize=4.0in\epsfbox{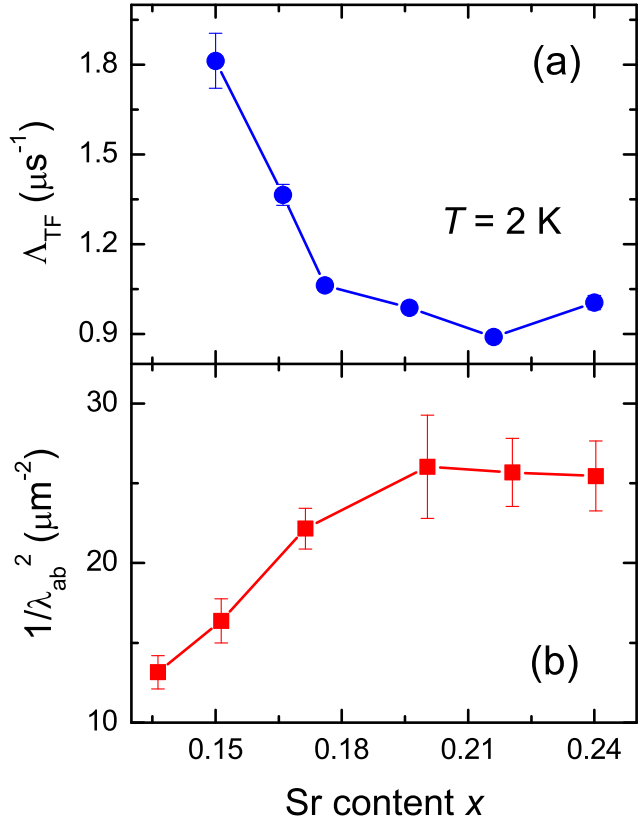}}
\caption{(a) The dependence of the muon spin depolarization rate $\Lambda$ 
in La$_{2-x}$Sr$_x$CuO$_4$ on Sr content $x$ (equivalent to hole-doping concentration)
at $T \! = \! 2$~K and $H \! = \! 7$~T. (b) The dependence of $\lambda_{ab}^{-2}$
on Sr content $x$ measured by ac susceptibility \cite{Panagopoulos:03}}
\label{figLSCOlowT}
\end{figure}

High TF-$\mu$SR measurements on La$_{2-x}$Sr$_x$CuO$_4$ at $H \! = \! 7$~T show
a complete breakdown of the low-field Uemura relation. 
As shown in figure~\ref{figLSCOlowT},
the depolarization rate $\Lambda$ does not track $\lambda_{ab}^{-2}$ 
below $x \! \sim \! 0.19$ --- and consequently $\Lambda$ does not scale with $T_c$
in the underdoped regime. The departure from Uemura scaling at high field is
apparently caused by the slowing down of electronic moments, which fluctuate 
too fast at $H \! = \! 0$ to be detected by ZF-$\mu$SR.
As mentioned earlier, LF-$\mu$SR measurements on La$_{2-x}$Sr$_x$CuO$_4$
show no evidence for spin fluctuations within the limits of detection 
\cite{Savici:05,MacDougall:06}.
Hence according to equation~(\ref{Knight}), the $\Lambda \! \propto \!  H$ 
behaviour observed just below and above $T_c$ is consistent with broadening of the 
local field distribution by heterogeneous static magnetism.
In addition, the gradual decrease of 
$\Lambda$ with increasing temperature well above the vortex 
solid-to-liquid transition (see figures~\ref{fig6_TR} and \ref{fig7_TR}), 
is consistent with a gradual increase of the fluctuation rate of 
the electronic moments. A likely source of the line broadening
is residual antiferromagnetic (AF) correlations from the parent insulator
La$_2$CuO$_4$ or static SDW order \cite{Hinkov:08}. 
At $H \! = \! 0$, neutron scattering measurements show that AF 
correlations persist to $x \! \sim \! 0.30$ \cite{Wakimoto:07}. While AF correlations
fluctuate on a shorter time scale than the $\mu$SR time window beyond $x \! \sim \! 0.12$ 
\cite{Weidinger:89,Niedermayer:98,Panagopoulos:02}, the fluctuations 
appear to be stabilized at high field.
 
As emphasized by MacDougall \etal \cite{MacDougall:10}, the field-induced
line broadening in La$_{2-x}$Sr$_x$CuO$_4$ 
is not accompanied by a sizeable shift of the average
internal field. This rules out commensurate long-range order or a spatially
uniform response of the sample to the external field. In other words, the
high TF-$\mu$SR measurements above $x \! = \! 0.12$ are consistent with
field-induced static AF order occurring in a volume fraction of the sample 
that decreases with increasing $x$. Above $x \! \sim \! 0.19$ the associated
line broadening is diminished to the point that the field inhomogeneity created
by the vortex lattice dominates. Consequently, $\Lambda$ tracks $\lambda_{ab}^{-2}$.
The observed saturation of $\lambda_{ab}^{-2}$ is consistent
with specific heat measurements in the superconducting state that
indicate an increasing number of unpaired
electrons beyond $x \! \sim \! 0.19$ \cite{Wang:07}, and
magnetization measurements signifying a reduced superconducting volume 
fraction in the heavily overdoped regime of La$_{2-x}$Sr$_x$CuO$_4$ \cite{Tanabe:05},
and other cuprates \cite{Wen:02}.
Furthermore, low TF-$\mu$SR measurements on cuprates show a reduction of the
superfluid density as the materials become progressively overdoped beyond 
$p \! \sim \! 0.19$ \cite{Uemura:93,Niedermayer:93,Bernhard:01}. 
Together these experiments indicate that 
heavily overdoped cuprates are microscopically phase separated into hole-rich and 
hole-poor regions \cite{Uemura:01}.

\begin{figure}
\centerline{\epsfxsize=5.0in\epsfbox{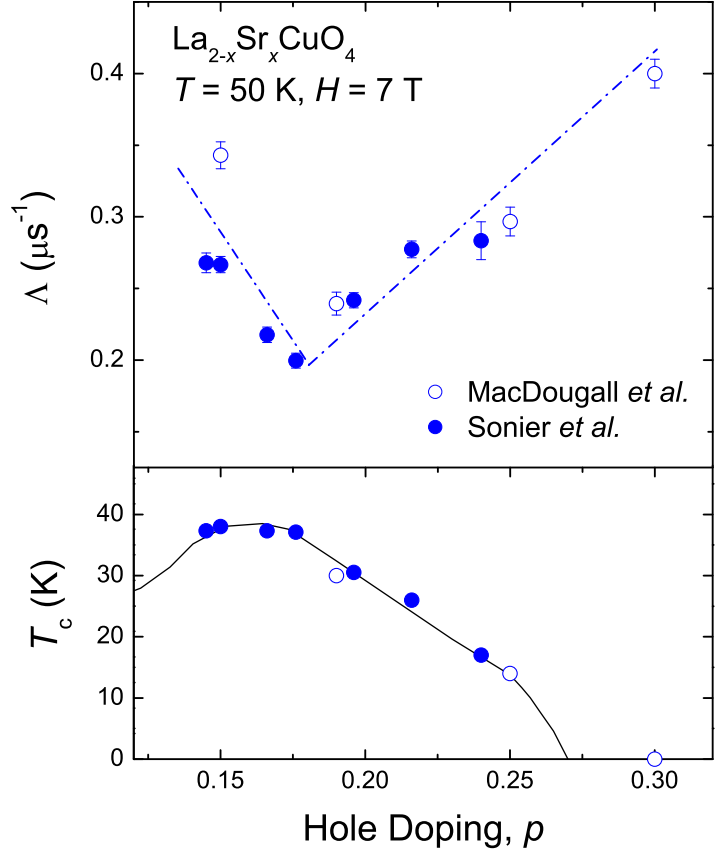}}
\caption{Hole doping dependences of $\Lambda$ in La$_{2-x}$Sr$_x$CuO$_4$  
at $T \! = \! 50$~K and $H \! = \! 7$~T (top panel), and the zero-field value of $T_c$ 
(bottom panel). The open circles are data from \cite{MacDougall:10} (denoted $\lambda$
in the original paper). The solid circles are data from \cite{Sonier:08,Sonier:10}}
\label{figLSCOaboveTc}
\end{figure}

With increasing temperature, fluctuations reduce the line broadening
associated with the AF correlations, and above $T_c$ a new trend in the doping
dependence of the line width emerges (see figure~\ref{figLSCOaboveTc}). 
Above $x \! \sim \! 0.19$, $\Lambda$
is observed to increase with hole doping \cite{MacDougall:10,Sonier:10}.
As first reported by MacDougall \etal \cite{MacDougall:06}, the width
of the local field distribution above $T_c$ continues to grow into
the heavily-overdoped {\it non-superconducting} regime.  
The increase of $\Lambda$ with $x$ above $x \! \sim \! 0.19$ is concomitant 
with the onset of an anomalous Curie term in 
normal-state bulk dc magnetization measurements \cite{Oda:91,Nakano:94,Wakimoto:05}.
The origin of the Curie-like paramagnetism has been somewhat of a mystery.
It has been proposed that phase separation and 
paramagnetism result from doping of holes into the Cu-3{\it d} 
orbital of the CuO$_2$ layers, rather than in the O-2{\it p} orbital 
\cite{Tanabe:05,Wakimoto:05}.
Doing so creates free Cu spins directly and/or by destroying the AF correlations 
between existing Cu spins. However, this explanation is speculative and
ongoing $\mu$SR studies are raising other possibilities. 

\begin{figure}
\centerline{\epsfxsize=6.5in\epsfbox{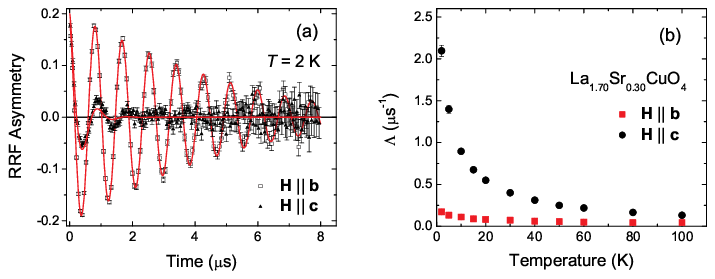}}
\caption{(a) TF-$\mu$SR asymmetry spectra of La$_{1.70}$Sr$_{0.30}$CuO$_4$
at $T \! = \! 2$~K (plotted in a rotating reference frame), for a field
of $H \! = \! 3$~T applied parallel to the $c$-axis, and parallel 
to the $b$-axis. (b) The temperature dependence of the depolarization
rate $\Lambda$ for the two different field orientations 
(from \cite{MacDougall:10}).}
\label{figMacDougallPRB10}
\end{figure}

MacDougall \etal \cite{MacDougall:10} have observed a highly anisotropic 
magnetic response in high TF-$\mu$SR measurements on La$_{1.70}$Sr$_{0.30}$CuO$_4$. 
As shown in figure~\ref{figMacDougallPRB10}, the field-induced line broadening is
significantly greater for a field applied parallel to the $c$-axis compared
to the line width for a field along the $b$-axis.
The anisotropy of $\Lambda$ bears some resemblance to the weaker anisotropy 
observed in La$_{1.875}$Ba$_{0.125}$CuO$_4$ above $T_c$ at $H \! = \! 6$~T 
by Savici \etal \cite{Savici:05}. The anisotropy of the magnetic response
in La$_{1.875}$Ba$_{0.125}$CuO$_4$ is almost certainly of the same origin 
as the anisotropy observed in bulk magnetic susceptibility measurements 
on La$_{2-x}$Sr$_x$CuO$_4$ below $x \! = \! 0.19$ \cite{Terasaki:92,Lavrov:01}.
The latter is associated with the effective susceptibility of residual
AF correlations from the parent insulator \cite{Johnston:89}. Silva Neto \etal
\cite{Neto:06} attribute the anisotropy of the magnetic susceptibility in the
parent compound La$_2$CuO$_4$ to the inclusion of Dzyaloshinskii-Moriya and
XY interactions in the 2D quantum Heisenberg antiferromagnet.
However, it is important to stress that the width of the internal field
distribution observed by high TF-$\mu$SR increases with doping above
$x \! \sim \! 0.19$, which is contrary to the diminishing contribution 
from residual AF correlations of the parent insulator. 
The high magnetic anisotropy and large value of the 
depolarization rate $\Lambda$ (for {\bf H}$\parallel${\bf c}) 
observed in heavily overdoped La$_{1.70}$Sr$_{0.30}$CuO$_4$ has prompted
MacDougall \etal \cite{MacDougall:10} to propose that the enhanced
line broadening above $x \! \sim \! 0.19$ is caused by a field-induced
staggered magnetization localized about the overdoped Sr ions.
However, the local spin structure cannot be directly obtained from
the TF-$\mu$SR experiments, and hence neutron scattering measurements in
an applied field are needed to confirm this hypothesis.  
 
Kopp \etal \cite{Kopp:07} have proprosed that ferromagnetic fluctuations
compete with superconductivity in the overdoped regime of cuprates, and that
a true ferromagnetic (FM) phase exists at $T \! = \! 0$ immediately beyond the
termination of the superconducting dome. Recent ZF-$\mu$SR measurements on
heavily overdoped non-superconducting
La$_{1.67}$Sr$_{0.33}$CuO$_4$ single crystals show the occurrence of a
frozen magnetic state below $T \! \sim \! 0.9$~K \cite{Sonier:10}. 
The measurements rule out the occurrence
of long-range FM order, but are in accord with dilute frozen electronic
moments or short-range magnetic order concentrated in dilute clusters. 
Consistent with the general idea of competing ferromagnetism, electronic band 
calculations by Barbiellini and Jarlborg \cite{Barbiellini:08} show that weak 
FM order develops about Sr-rich clusters in the heavily overdoped regime. 
Whether the magnetism detected by ZF-$\mu$SR is in the form of FM clusters 
remains to be determined. Moreover, it is unclear at this time how the
field-induced magnetism above $x \! \sim \! 0.19$ is related to the magnetism
detected in La$_{1.67}$Sr$_{0.33}$CuO$_4$ at $H \! = \! 0$. Interestingly,
the strong magnetic anisotropy observed by MacDougall \etal \cite{MacDougall:10}
in La$_{1.70}$Sr$_{0.30}$CuO$_4$ and by others in lower-doped samples 
is absent in La$_{1.67}$Sr$_{0.33}$CuO$_4$ \cite{Sonier:10}.  

\subsection{YBa$_2$Cu$_3$O$_y$: Evidence for inhomogeneous superconducting fluctuations above $T_c$}

Although the heavily overdoped regime cannot be reached in YBa$_2$Cu$_3$O$_y$ 
by oxygen doping, static magnetism at $H \! = \! 0$  \cite{Kiefl:89,Sanna:04,Miller:06}
occurs at a lower hole-doping concentration than in La$_{2-x}$Sr$_x$CuO$_4$.
Furthermore, the maximum field of $H \! = \! 7$~T in a TF-$\mu$SR experiment 
is not high enough to induce long-range SDW order in YBa$_2$Cu$_3$O$_y$.
Consequently, the width of the internal field distribution can be investigated
over a wide range of doping where field-induced static magnetism 
does not necessarily dominate the muon depolarization rate.     

\begin{figure}
\centerline{\epsfxsize=3.5in\epsfbox{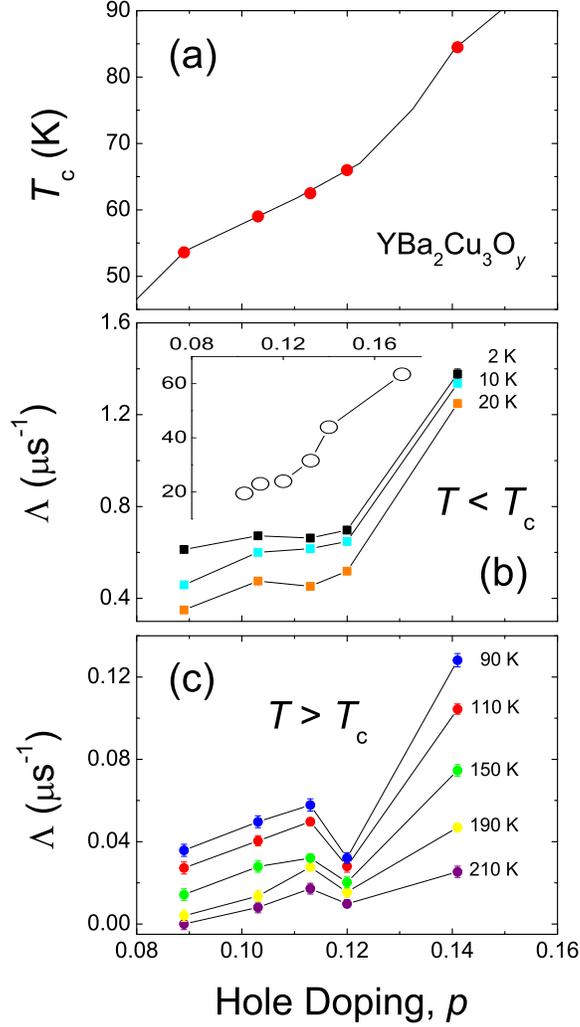}}
\caption{Hole doping dependences of (a) $T_c$ at $H \! = \! 0$, 
(b) $\Lambda$ below $T_c$ at $H \! = \! 7$~T, and  
(c) $\Lambda$ above $T_c$ at $H \! = \! 7$~T for YBa$_2$Cu$_3$O$_y$
(from \cite{Sonier:08}).
The inset of (b) shows the hole doping dependence of $1/\lambda_{ab}^2$ (in $\mu$m$^{-2}$) 
in YBa$_2$Cu$_3$O$_y$ determined by a proper analysis of the internal field distribution 
of the vortex solid phase below $T_c$.}
\label{figYBCOhole}
\end{figure}

The hole-doping dependence of $\Lambda$ in the underdoped regime of YBa$_2$Cu$_3$O$_y$ 
has been measured at $H \! = \! 7$~T by TF-$\mu$SR \cite{Sonier:08} in
samples that do not exhibit static magnetism at $H \! = \! 0$.
As shown in figure~\ref{figYBCOhole}(b), $\Lambda$ tracks $T_c$ at 
temperatures below $T_c$, in agreement with the ``Uemura plot'' deduced at lower
field \cite{Uemura:89}. Thus in stark contrast to underdoped La$_{2-x}$Sr$_x$CuO$_4$, 
the dominant contribution to $\Lambda$ at high field remains 
the static field inhomogeneity created by vortices. 
Indeed the doping dependence of $\Lambda$ below $T_c$ closely
resembles the doping dependence of $\lambda_{ab}^{-2}$ \cite{Sonier:07b}.
What is most surprising is that the doping dependence of $\Lambda$ continues to follow 
$T_c$ and $\lambda_{ab}^{-2}$ at temperatures above $T_c$ 
(as shown in figure~\ref{figYBCOhole}(c)).
Moreover, the subtle suppression of $\Lambda$ that is observed below $T_c$
near a hole-doping concentration of $p \! = \! 1/8$ becomes more pronounced 
above $T_c$.  

In zero field, the rms deviation of the local magnetic field 
at the muon site increases as the hole doping concentration $p$ is increased
above $p = 0$ \cite{Kiefl:89,Niedermayer:98}. This is due to the gradual 
destruction of the AF phase of the parent compound. 
However, in both YBa$_2$Cu$_3$O$_y$ and La$_{2-x}$Sr$_x$CuO$_4$, the rms 
deviation of the local field vanishes deep in the underdoped regime. 
Moreover, in the case of La$_{2-x}$Sr$_x$CuO$_4$
there is enhanced spin freezing near 1/8 hole doping \cite{Julien:03}, which 
is believed to be associated with hole and spin ``stripe'' ordering \cite{Kivelson:03}.
While this is not observed in zero-field measurements on YBa$_2$Cu$_3$O$_y$, there is a 
suppression of $T_c$ in the neighbourhood of $p = 1/8$ \cite{Liang:06}.   
Furthermore, Zn-doped YBa$_2$Cu$_3$O$_y$ shows an enhancement of the width of the
local field distribution at $p \! \sim \! 1/8$,
suggesting that there are dynamic stripes that are pinned by the Zn impurity 
\cite{Akoshima:00}. An applied magnetic field appears to have the same effect,
because the levelling off of $\Lambda$ near $p \! = \! 1/8$ in 
figure~\ref{figYBCOhole}(b) is consistent with a suppression of the superfluid density 
due to competing stripe correlations. With increasing temperature the 
line width associated with the vortices decreases, and above $T_c$ the magnetic
component of the stripe order should cause an enhancement, rather than the observed
suppression of $\Lambda$ near $p \! = \! 1/8$.
This behaviour and the larger value of $\Lambda$ at $p \! = \! 0.141$ 
rules out SDW order or remnant Cu spins of the AF phase as the primary
source of the field-induced line broadening above $T_c$.

It is clear that $\Lambda$ above $T_c$ tracks the bulk superconductivity of
underdoped YBa$_2$Cu$_3$O$_y$ below $T_c$. One possible explanation is that 
$\Lambda$ reflects the density of the normal state charge carriers. In other words, 
the applied magnetic field affects the spins of the unpaired 
electrons that eventually bind to form phase-coherent Cooper pairs below $T_c$. 
To have an effect on the muon depolarization rate, these electrons must localize.
Yet there is no experimental evidence to support this scenario.
Low-temperature thermal conductivity measurements in an external field 
by Sun \etal \cite{Sun:04} suggest that quasiparticle localization occurs,
but only for $y \! \leq \! 6.50$ and not necessarily above $T_c$. 
In fact Sutherland \etal \cite{Sutherland:05} have
argued that such localization does not occur even below $y \! = \! 6.50$,
and that the observed suppression of the thermal conductivity 
in a field can be explained by scattering of quasiparticles from the vortex cores.

Sanning tunneling microscopy STM measurements on 
Bi$_2$Sr$_2$CaCu$_2$O$_{8 + \delta}$ by Gomes \etal \cite{Gomes:07} indicate that
superconducting-like energy gaps exist in random nanometer-size patches well
above $T_c$, and that these regions proliferate as the sample is cooled through 
$T_c$. Assuming such spatial inhomogeneity is universal to the cuprates,
the contribution of superconducting fluctuations to the Nernst signal above
$T_c$ \cite{Xu:00,Wang:01,Wang:06,Cyr:09,Daou:09} may be attributed to
a vortex liquid residing in superconducting droplets above $T_c$. 
Since the doping dependence of $\Lambda$ in YBa$_2$Cu$_3$O$_y$ above $T_c$ 
resembles the vortex-like Nernst signal \cite{Daou:09}, it is tempting to 
directly associate the field-induced line broadening with the occurrence of
a vortex liquid. Since TF-$\mu$SR studies of the vortex-liquid phase of cuprates 
below $T_c$ show a severely motionally-narrowed line width that decreases with
increased field \cite{Sonier:00}, the break up into superconducting patches
appears to be the key to understanding the field-induced
line broadening in YBa$_2$Cu$_3$O$_y$ above $T_c$. The muons will sense
a unique time-averaged local diamagnetic field associated with each nanometer-size
patch exhibiting fluctuating superconductivity. A distribution of such
time-averaged local fields $\delta \langle B(t) \rangle$ broadens the $\mu$SR 
line width according to equation~(\ref{Knight}). In this picture,
each superconducting patch is characterized by a distinct local $T_c$ 
that is higher than the bulk value of $T_c$. As the temperature is raised, an
increasing number of the superconducting patches turn normal, and consequently 
$\delta \langle B(t) \rangle$ is reduced. This is consistent with the
observed reduction of $\Lambda$ in YBa$_2$Cu$_3$O$_y$ with increasing temperature.

Torque magnetometry measurements on highly-overdoped
Tl$_2$Ba$_2$CuO$_{6+\delta}$ by Bergemann \etal \cite{Bergemann:98}
provide insight into the field dependence of $\Lambda$ at temperatures above $T_c$.
A linear bulk diamagnetic response is observed that
persists to temperatures well above $T_c$, which is contrary to the field
dependence of a vortex liquid. This unusual result has been attributed to
inhomogeneous superconductivity, whereby small regions (tens of nanometers in
size) with local $T_c$'s much higher than the bulk $T_c$ each exhibit a 
linear diamagnetic response \cite{Geshkenbein:98}. Since the TF-$\mu$SR
line width is proportional to an inhomogeneous distribution of time-averaged
local diamagnetic susceptibilities, this model explains why 
$\Lambda$ increases with field.

\subsection{Further considerations}

One may wonder why there has not been widespread reports of NMR experiments
detecting a similar inhomogeneous magnetic field response in cuprates above $T_c$. 
Because the muon is a spin 1/2 particle, it is a pure local magnetic probe. 
In contrast, the important NMR active nulcei in the cuprates have spin greater than 
1/2, and hence in addition to a magnetic dipole moment they possess 
an electric quadrupole moment that also couples to the local environment.
Consequently, any contribution of static field inhomogeneity to the NMR line width
must be separated from broadening due to electric field gradients.
 
Haase \etal \cite{Haase:00,Haase:02} have shown that there is 
magnetic broadening of the $^{63}$Cu and $^{17}$O NMR line widths 
in La$_{2-x}$Sr$_x$CuO$_4$ above $T_c$. Like $\Lambda$ in the $\mu$SR experiments, the NMR 
line widths in La$_{2-x}$Sr$_x$CuO$_4$ have a Curie-like dependence on temperature 
and are proportional to the applied field. The magnetic broadening of the
NMR line width is explained by static inhomogeneity on the scale of a few lattice
constants, which has been 
associated with non-uniform polarization of the Cu electronic moments
by the external field. Furthermore, the NMR experiments by Haase \etal 
reveal that the static field inhomogeneity is correlated with spatial
modulations of the electric field gradients. However, the occurrence of
static inhomogeneity above $T_c$, unrelated to sample
quality ({\it i.e.} variations in local doping), has yet to be 
established by NMR in other cuprate systems. This may indicate that the
normal-state $\mu$SR line width is associated with slowly fluctuating magnetism that is
outside the NMR time window. But a more likely explanantion is that the
regions contributing to the tails of the local field distribution are too dilute 
to be detected by NMR. Here $\mu$SR has a distinct advantage. The nearly 
100~\% initial spin 
polarization and the large gyromagnetic ratio of the $\mu^+$ makes $\mu$SR
sensitive to regions of small volume fraction.   
In a dilute system the local field distribution has a Lorentzian shape,
and even in the static limit this causes exponential relaxation of the TF-$\mu$SR 
signal. While a pure Lorentzian lineshape is unaffected by motional averaging
\cite{Silsbee:83}, in the sample there is always a maximum local field. 
Thus $\Lambda$ can be reduced by fluctuating internal fields, as well as a 
a gradual disappearance of magnetic regions.

\section{Conclusions}

It has been explained here that the dominant contribution to the 
inhomogeneous field distribution detected in underdoped YBa$_2$Cu$_3$O$_y$ at
high field above $T_c$ appears to be time-averaged fields from fluctuating 
diamagnetic regions. The high TF-$\mu$SR measurements indicate that superconducting
correlations persist to temperatures significantly above the region where
a Nernst effect due to superconducting fluctuations is observed.
But the inhomogeneous superconducting fluctuations detected in
YBa$_2$Cu$_3$O$_y$ by TF-$\mu$SR are uncorrelated with the pseudogap
temperature $T^*$, and even persist above $T^*$. It is important to stress
that the sensitivity of the TF-$\mu$SR line width to the fluctuating diamagnetism 
necessarily implies spatial inhomogeneity.

By contrast, the effect of superconducting fluctuations on the TF-$\mu$SR
depolarization rate of La$_{2-x}$Sr$_x$CuO$_4$ above $T_c$ is masked by 
field-induced spin magnetism. A crossover in behaviour near $x \! \sim \! 0.19$
distinguishes remnant AF correlations of the parent insulator or SDW order
from an increasing 
heterogeneous magnetic response in the heavily overdoped regime. The source
of the heterogeneous magnetism above $x \! \sim \! 0.19$ is unknown, but its 
appearance in the heavily overdoped regime may explain the demise and 
eventual termination of superconductivity at high doping. 
        
\ack
I would like to thank G M Luke, G J MacDougall and A T Savici for sharing and
discussing their $\mu$SR data. I would also like to thank 
W N Hardy, D A Bonn, R Liang, Y Ando, S Komiya, W A Atkinson, V P Pacradouni, 
C V Kaiser, and S A Sabok-Sayr for their contributions to 
some of the $\mu$SR work discussed here. The writing of this article 
was supported by the Natural 
Science and Engineering Research Council of Canada, and the Quantum Materials 
program of the Canadian Institute for Advanced Research.

\end{document}